# Rayleigh-Darcy convection in a porous layer:
# A comparison of near-critical and normal fluid phases

E. B. Soboleva

*Institute for Problems in Mechanics, Russian Academy of Sciences,
prospect Vernadskogo, 101, b.1,
119526 Moscow, Russia*

Email address: soboleva@ipmnet.ru

Gravity-driven convection of fluid at parameters near its thermodynamic critical point inside a porous layer heated from below (Rayleigh-Darcy convection) is studied. The fluid having the temperature slightly above the critical one is one-component. The hydrodynamic model describing a high compressible fluid phase at variable physical properties inside a solid matrix at a uniform porosity is analyzed. A near-critical fluid is assumed to be the van der Waals gas. In the limit of small variations in the density and thermodynamic coefficients where the Oberbeck-Boussinesq approximation is applicable, the correlation relations for the key criteria of similarity under the stratification effect are obtained. These relations connect the model parameters (the Rayleigh-Darcy and Prandtl numbers appearing in the dimensionless governing equations) with the criteria of similarity (the real Rayleigh-Darcy and Prandtl numbers characterizing convection actually). Steady-state Rayleigh-Darcy convection near the stability threshold in conditions of the Oberbeck-Boussinesq approximation to be valid is simulated numerically. A set of solutions is analyzed with the use of the correlation relations resulting in a universal dependency of the corrected Nusselt number on the real Rayleigh-Darcy number described by a single curve. Convection of near-critical fluid is compared with that of normal fluid. The temperature difference at boundaries corresponding to the convection onset and depending on the Schwarzschild and Rayleigh criteria is analyzed analytically. As obtained, near the critical point, the threshold temperature difference at boundaries becomes independent of solid matrix and is determined solely by the critical adiabatic temperature gradient of fluid phase.

## I. INTRODUCTION

Fluids at parameters near the gas-liquid critical point are of great interest because of their abnormal physical properties leading to peculiarities in hydrodynamic behavior. We consider here only macroscopically homogeneous one-component fluids at the temperature slightly above the critical one. A very high compressibility of such fluids is responsible for a strong thermo-mechanical coupling which, in unsteady regimes, gives rises to adiabatic heating (heating due to compression) called the piston effect [1-4]. In steady regimes, a high compressibility accounts for hydrostatic effects which become not negligible in laboratory scales [5]. Our attention focuses on stability problems in fluids subjected to the density stratification, the onset of gravity-driven convection and dynamic behavior inside a horizontal layer heated from below (in the case of a pure fluid, the Rayleigh-Bénard convection). In stratified fluids, the adiabatic temperature gradient plays a key role [6] and the onset of convection is controlled by the Schwarzschild criterion. The stabilizing mechanism in this case is related to an adiabatic process when an elementary volume of fluid rises, expands and is cooled. If the temperature gradient in the fluid is less then the adiabatic temperature gradient, a convective motion does not develop. In incompressible fluids, the onset of convection is controlled by the Rayleigh criterion that is determined by the viscous and thermal dissipation. The onset of convection in near-critical fluids involving two mentioned mechanisms was firstly analysed in [7] and, later, in [8]; the piston effect does not influence the onset of convection [9]. Unsteady problems are associated mostly with the interplay between convection and the piston effect [10, 11], oscillatory behavior under the applied constant heat current [12-15], and reverse transition to stability through the Schwarzschild line [16-18]. Steady regimes reaching in a long time are considered to study the reduction in heat exchange due to stratification [19], the stabilizing effect of the adiabatic temperature gradient [20], and universal features in the heat transfer of near-critical fluids

[13, 21]. Notice that the experiments on convective heat transport in near-critical $^3$He [22] triggered series of numerical studies of convection in high compressible fluids [12-15, 19, 21]. The steady-state results of these experiments and simulations showed that the convection heat current versus the reduced Rayleigh number collapses onto a single curve [13]. The similar universal curve was drawn in [21] resulted from simulations of near-critical Rayleigh-Bénard convection.

In this paper, the object of our study is a near-critical fluid inside a porous matrix. These systems are of great interest for many industrial applications in the areas of chemical and materials engineering, extractions, microparticle processing and impregnation. The behaviour of such systems is complicated and subjected to finite-size effects. As found experimentally [23, 24], a critical scaling behaviour of fluid phase can be suppressed because the correlation length of density fluctuations does not exceed the pore size. As reported in [25, 26], a shift of critical temperature and density due to the matrix-fluid interaction can occur. However, we do not take into account such distinct critical phenomena which are significant mostly in nano- and micropores. Investigations of hydrodynamic behaviour of confined near-critical fluids are few in number. Numerical simulations of convection in near-critical porous media based on the classical Oberbeck-Boussinesq approximation were carried out [27, 28]. Later, a linear asymptotic analysis of the piston effect in the near-critical van der Waals gas inside a porous layer in one-dimensional approximation was performed [29]. The piston effect and unsteady convective flows in rectangular porous cavities in two-dimensional approximation were simulated numerically [30, 31].

In this paper, gravity-driven convection in a near-critical fluid saturating a porous layer heated from below (Rayleigh-Darcy convection) is studied numerically in two-dimensional geometry on the basis of the full mathematical model developed in [30, 31]; a near-critical fluid is described as the van der Waals gas. The theoretical background is given in Sec. II. In the limit of small variations in the density and thermodynamic coefficients where the full model is restricted to the Oberbeck-Boussinesq approximation, the correlation relations for the key criteria of similarity under the stratification effect are obtained. These relations connect the model Rayleigh-Darcy and Prandtl numbers appearing in the governing equations with the real Rayleigh-Darcy and Prandtl numbers characterising convection actually and being the criteria of similarity. In Sec. III, steady-state Rayleigh-Darcy convection slightly above the stability threshold in conditions of the Oberbeck-Boussinesq approximation to be applicable is simulated numerically. The set of solutions is analyzed with the use of the correlation relations resulting in a universal dependency of the corrected Nusselt number on the real Rayleigh-Darcy number described by a single curve. This dependency agrees with the analytical solution for a low compressible fluid phase. Convection of near-critical fluid is compared with that of normal fluid at the equal criteria of similarity to realise the effect of high compressibility in the fist case. In Sec. IV, some parameters corresponding to the convection onset depending on the Schwarzschild and Rayleigh criteria are analyzed analytically. Results and conclusions are given in Sec. V.

## II. THEORETICAL BACKGROUND

### A. Full mathematical model and numerical code

A two-phase system composed of porous matrix and one-component near-critical fluid is considered. The matrix is isotropic and characterised by the porosity $\varphi$ and permeability $K'$ to be constants. The dimensional values are marked by a prime whereas the nondimensional values have no prime. The resistance of matrix is a linear function of fluid velocity. To describe critical properties of fluid phase, the equation of state yielding the discrepancy of the specific heat at constant pressure and compressibility at the critical point is used. The specific heat at constant volume is assumed to be constant. The thermal conductivity of fluid $\lambda'_f$ diverges at the critical point as a power-law function: $\lambda'_f = \lambda'_0 (1 + \Lambda((T'-T'_c)/T'_c)^{-\psi})$. Here $T'$ is the temperature and $\lambda'_0$, $\Lambda$, $\psi$ are the constants; subscript "$c$" corresponds to the critical point. The phases are assumed to be in local thermal equilibrium that is the solid and fluid have identical temperatures.

We use the procedure of two-scale splitting of the pressure in the governing equations. The pressure $P'$ is replaced with the sum of volume-average component $\langle P \rangle'$ and rest component $p'$ associated with dynamic changes and stratification: $P' = \langle P \rangle' + p'$. The component $\langle P \rangle'$ does not depend on space variables and can be excluded from the pressure gradient, that is $\vec{\nabla} P' = \vec{\nabla} p'$. In low-speed flows, $p' \ll P'$ therefore the round-off error of calculation of $\vec{\nabla} p'$ is several orders of magnitude smaller than that of $\vec{\nabla} P'$. The pressure components are normalized by different scales corresponding to their characteristic changes: $P'$ and $\langle P \rangle'$ by $B' \rho'_c T'_c$ whereas $p'$ by $\rho'_c U'^2$. Here, $\rho'_c$ and $U'$ are the critical density of fluid and magnitude of filtration velocity, and $B' = R'/\mu'_g$ is the constant; $R'$ is the universal gas constant, and $\mu'_g$ is the molecular weight. The pressure decomposition in nondimensional form is determined by the first equation Eq. (5) below. Note that splitting of the pressure and further transformations in the governing equations do not lead to the "acoustic filtration". We do not remove the component $p$ from the equation of state and consider the full equation determined by the general relation Eq. (4) below. The "acoustic filtration" will take place if the full pressure $P$ in the equation of state replaces with $\langle P \rangle$ not feeling changes due to motions that is if the equation $\langle P \rangle = \langle P \rangle(\rho, T)$ is used instead of Eq. (4). The present model can give acoustic solutions as well. However when the characteristic time of problem becomes much longer than the acoustic time, the local contribution from sonic waves goes to zero. In this limit, the full model with two-scale splitting of the pressure approaches to the restricted model with "acoustic filtration" [32]. Our consideration is based on the model [33] employed firstly to the ideal gas. The pressure decomposition in [33] is close in a sense to the multiple pressure variables method (MPV-method) which is successively developed to compute low Mach number flows (see [34] and references therein).

The mathematical model includes the continuity, momentum, energy equations and the two-parametric equation of state. Two equations associated with the pressure decomposition are added. The governing equations can be written in the following dimensionless form [30, 31]:

$$\varphi \frac{\partial \rho}{\partial t} + \nabla \cdot (\rho \vec{U}) = 0 \quad (1)$$

$$\frac{\rho}{\varphi} \frac{\partial \vec{U}}{\partial t} = -\vec{\nabla} p + \frac{1}{Fr} \rho \vec{g} - \frac{1}{Da \, Re} \frac{1}{K} \vec{U} \quad (2)$$

$$((1-\varphi)\rho_s c_s + \varphi \rho) \frac{\partial T}{\partial t} + \rho (\vec{U} \cdot \nabla) T =$$

$$= -(\gamma_0 - 1) T \left( \frac{\partial P}{\partial T} \right)_\rho \nabla \vec{U} + \frac{\gamma_0}{Re \, Pr_0} \nabla \cdot (\lambda_m \nabla T),$$

$$\lambda_m = 1 + \varphi \lambda_0 \Lambda (T-1)^{-\psi} \quad (3)$$

$$P = P(\rho, T) \quad (4)$$

$$P = \langle P \rangle + \gamma_0 M^2 p, \quad \int_\Omega p \, d\varpi = 0 \quad (5)$$

Here, $\rho$, $\vec{U}$, $\vec{g}$, $\rho_s$, $c_s$, $d\varpi$ and $\Omega$ are the density of fluid, filtration velocity, mass force acceleration, density and specific heat of solid matter, element of volume and total considered volume. The scales are: height $h'$, velocity $U'$, time $h'/U'$, mass force acceleration $g'$, critical parameters $\rho'_c$ and $T'_c$, viscosity and specific heat at constant volume of fluid $\eta'_0$ and $c'_{v0}$, permeability $K'_0$, thermal conductivity of porous medium $\lambda'_{m0}$: $\lambda'_{m0} = (1-\varphi)\lambda'_s + \varphi \lambda'_{f0}$. Subscripts "$s$", "$f$", "$m$", and "$0$" correspond to the solid, fluid phases, medium "solid+fluid", and reference values (being far from the critical point). The dimensionless parameters are the Froude, Prandtl, Reynolds, Mach, Darcy numbers, dimensionless permeability and ratio of specific heats of non-critical fluid

$$Fr = \frac{U'^2}{g'h'}, \quad Pr_0 = \frac{(c'_{v0} + B')\eta'_0}{\lambda'_{m0}},$$

$$Re = \frac{\rho'_c U' h'}{\eta'_0}, \quad M = \frac{U'}{\sqrt{\gamma_0 B' T'_c}}, \quad Da = \frac{K'_0}{h'^2},$$

$$K = \frac{K'}{K'_0}, \quad \gamma_0 = 1 + \frac{B'}{c'_{v0}} \quad (6)$$

The energy equation Eq. (3) written in dimensional variables includes the specific heat at constant volume of fluid $c_v'$ in its left-hand side. Following from the terminology in [20], this form of energy equation is referred to the "$c_v$ formalism".

As mentioned above, the value $c_v'$ does not feel criticality and stays constant in our consideration. We take $c_v' = c_{v0}'$. On some transformations, Eq. (3) may be written in another form including the specific heat at constant pressure $c_p'$ and known as "$c_p$ formalism"; this nondimensional form is

$$\left((1-\varphi)\rho_s c_s + \varphi \rho c_p\right)\frac{\partial T}{\partial t} + \varphi \rho c_p (\vec{U} \cdot \nabla)T =$$
$$= -(\gamma_0 - 1)\varphi \frac{T}{\rho}\left(\frac{\partial \rho}{\partial T}\right)_P \frac{dP}{dt} + \frac{\gamma_0}{\text{RePr}_0}\nabla \cdot (\lambda_m \nabla T) \quad (7)$$

Near the critical point, the specific heat $c_p$ grows unboundedly leading to some difficulties in designing appropriate numerical algorithms for Eq. (7); a precision of numerical solutions at $c_p \to \infty$ may be lost. Avoiding these difficulties we designed a numerical solver based on the energy equation in the form of Eq. (3). But, as Eqs. (3) and (7) are equivalent to each other, we shall use the second one in an analytical analysis below.

The density and pressure stratification of fluid phase due to the gravity force is described in isothermal conditions by linear functions of the position $\vec{r}$:

$$\rho_i = \rho^+\left(1 + \left(\frac{\partial \rho^+}{\partial P^+}\right)_{T^+}\frac{\gamma_0 M^2}{Fr}\vec{g}\cdot(\vec{r}-\vec{r}^+)\right) \quad (8)$$

$$p_i = p^+ + \rho^+ \frac{1}{Fr}\vec{g}\cdot(\vec{r}-\vec{r}^+) \quad (9)$$

The subscript "$i$" and superscript «+» correspond to the initial conditions and position of centre of calculation domain. The fluid slips at boundaries.

We use for a fluid phase the van der Waals equation of state

$$P = \frac{\rho T}{1 - a\rho} - b\rho^2, \quad a = 1/3, \quad b = 9/8 \quad (10)$$

giving the expressions of derivatives $(\partial P/\partial T)_\rho$, $(\partial \rho/\partial T)_P$, and $(\partial \rho^+/\partial P^+)_{T^+}$ in the governing equations. It is possible to incorporate different equations of state into the model above describing critical properties of fluid more exactly. For example, there are the crossover van der Waals equation giving accurate critical exponents and other equations (see [35] and references therein). However we do not use such complicated equations of state in the present stage of study.

We have designed a 2D numerical code for integrating the governing equations. To improve a precision of numerical solutions near the critical point, the variables $\rho$, $\langle P \rangle$, and $T$ in temporal and spatial derivatives substitute by the reduced density, pressure and temperature $m$, $\pi$, and $\varepsilon$, respectively, defined as $m = (\rho' - \rho'_c)/\rho'_c = \rho - 1$, $\pi = (\langle P' \rangle - P'_c)/(B'\rho'_c T'_c) = \langle P \rangle - 1/(1-b) + a$, and $\varepsilon = (T' - T'_c)/T'_c = T - 1$. Using variables $m$, $\pi$ and $\varepsilon$, the van der Waals equation of state Eq. (10) is transformed into the form

$$\pi = \frac{(1+m)\varepsilon}{1 - b(1+m)} + \frac{b^2 m^3}{(1-b)^3(1-b(1+m))} \quad (11)$$

The governing differential equations are discretizied by the finite difference method [36]. A staggered non-uniform rectangular grid in which velocity nodes are located in between nodes of dynamic pressure is used. The density and temperature variables are defined at the same grid points as the dynamic pressure. Spatial derivatives are approximated by second-order central difference schemes. Temporal derivatives are approximated by first-order forward difference formulae.

The discretization equations are solved successively in several steps. First, the volume-average pressure $\pi$ and density $m$ are calculated by an iterative method; $\pi$ does not depend on space whereas $m$ is represented by its values at grid points. In this step, the equation of state Eq. (11) written at grid points and the integral mass of fluid (the sum of mass over all control volumes) which should be retained are calculated alternatively. Iterations are carried out till the integral mass converges.

In the second step, the velocity $\vec{U}$ and dynamic pressure $p$ are calculated from the continuity and momentum equations using the SIMPLE-type algorithm [37]. The starting values of $\vec{U}$ and $p$ correspond to the time level denoted by the superscript "$n$". The momentum equation Eq. (2) which is a linear ordinary differential equation for $\vec{U}$ may be integrated analytically over the time interval $\tau$. The solution is

$$\vec{U}^* = \vec{U}^n \exp\left(-\frac{\varphi}{Da\operatorname{Re}K\rho}\tau\right)$$
$$+ Da\operatorname{Re}K\left(\nabla p^n - \frac{1}{Fr}\rho\vec{g}\right)\left(\exp\left(-\frac{\varphi}{Da\operatorname{Re}K\rho}\tau\right) - 1\right) \quad (12)$$

Equation (12) gives the intermediate velocity field $\vec{U}^*$ corresponding to the pressure field at the starting time level. The velocity and pressure fields should be corrected by the increments $\delta\vec{U}$ and $\delta p$ giving the values $\vec{U}^{**}$ and $p^*$

$$\vec{U}^{**} = \vec{U}^* + \delta\vec{U}, \quad p^* = p^n + f\delta p \quad (13)$$

where $f$ is the under-relaxation factor for pressure. We use $f = 0.5$ in our calculations. The magnitudes $\vec{U}^{**}$ and $p^*$ are related by the equation similar to Eq. (12) in which $\vec{U}^*$ is replaced with $\vec{U}^{**}$ and $p^n$ with $p^*$. Equation (12) is subtracted from the appropriate equation for $\vec{U}^{**}$ and $p^*$ leading to the relation

$$\delta\vec{U} = Da\operatorname{Re}K\left(\exp\left(-\frac{\varphi}{Da\operatorname{Re}K\rho}\tau\right) - 1\right)\nabla\delta p \quad (14)$$

The second condition on $\delta\vec{U}$ and $\delta p$ results from the correction of the continuity equation which should be satisfied at $\vec{U}^{**}$ and $p^*$. It has the form

$$\rho\nabla(\delta\vec{U}) + \delta\vec{U}\nabla(\rho) = -\varphi\frac{\partial\rho}{\partial t} - \rho\nabla(\vec{U}^*) - \vec{U}^*\nabla(\rho) \quad (15)$$

Affecting Eqs. (14), (15) by the discretization and combining the results, one can obtain the finite difference analog of the Poisson equation for $\delta p$. After some simple transformations, we lead to the linear algebraic equations for the grid-point values of $\delta p$ which are solved using the line Gauss-Seidel method. So, in this step, we calculate $\vec{U}^*$ by Eq. (12), $\delta p$ by the Poisson equation, $\delta\vec{U}$ by Eq. (14), $\vec{U}^{**}$ and $p^*$ by Eqs. (13). Corrections of $\vec{U}$ and $p$ are repeated till the converged solutions are obtained.

In the third step, the energy equation is integrated to give the temperature $\varepsilon$. Equation (3) is replaced with the finite difference equation. Solution of the algebraic equations is constructed by the line Gauss-Seidel method as well.

In the fourth step, the grid-point values of $\vec{U}$, $p$ and $\varepsilon$ are corrected explicitly by the discretization equations obtained early. At last, the grid-point values of $p$ are shifted by the same constant so that the second equation of Eqs. (5) is satisfied. This procedure leads to the shift in $\langle P \rangle$ and, consequently, in $\pi$ since the full pressure $P$ in the first equation of Eqs. (5) has to be retained. Because all the variables are calculated for not the same grid points, we define, if necessary, some values through linear interpolation.

Before all steps, boundary conditions are set. For the velocity vector components, the Dirichlet or Neumann boundary conditions are applied (the velocity component or its derivative is zero) depending on whether the grid points lie on the physical boundary of calculation domain or do not. For the dynamic pressure, the Neumann boundary conditions are applied setting the derivative to be zero. The temperature boundary conditions are defined by the considered problem. Conditions on the density are resulted from the condition on the boundary mass flux that should be zero.

A precision of numerical calculations is checked by the heat balance. The energy equation Eq. (5) is integrated in space and time giving the heat changes in the whole calculated domain from the initial time instant. We perform the numerical integration by summing the terms through the all control volumes and time levels. The resultant left-hand and right-hand sides those have to be equal to each other in an exact solution are compared. We obtained that, in typical simulations, the balance between the left-hand and right-hand sides is retained within 0.1 %.

Using the designed numerical code, simulations of the piston effect and steady-state gravity-driven convection in vertical porous layers were successively performed [31]. As obtained, the numerical results are in good agreement with analytical predictions. The present code is based on the early code employed to computations of convective problems in pure near-critical fluids [21, 38, 39] which was modified and extended to two-phase systems.

## B. Oberbeck-Boussinesq approximation and the correlation relations

We consider a fluid on the critical isochore saturating a horizontal porous layer at the width $h'$. Let the temperature at the top and bottom boundaries be $T'_i$ and $T'_i + \Theta'$. If variations in density and thermodynamic coefficients of fluid phase are small, the Oberbeck-Boussinesq approximation can be employed. Early, the Oberbeck-Boussinesq approximation was successively used as a model for numerical simulations of one-component near-critical dynamic phenomena [13, 20]. Two conditions for this approximation to be valid are the same as in a pure near-critical fluid [13]:

$$\Theta << \varepsilon, \quad \varepsilon^{\beta+\gamma} >> \Gamma'_a h'/T'_c \qquad (16)$$

Here, $\beta$, $\gamma$ are the critical exponents (in the van der Waals gas, $\beta = 1/3$, $\gamma = 1$), $\Gamma'_a$ is the adiabatic temperature gradient, and $\Theta = \Theta'/T'_c$.

The full model Eqs. (1)-(5) is transformed into the Oberbeck-Boussinesq approximation in the standard way. In this Subsection, we consider the energy equation in the form of Eq. (7) instead of Eq. (3) as "$c_p$ formalism" includes the separated term responsible for the piston effect (first term on the right-hand side) and is more convenient for analysing below. The equation of state is simplified to the linear form relating the density deviation from the initial value $\rho - \rho_i$ with the temperature one $T - T_i$: $\quad \rho - \rho_i = -\alpha_p \rho_i (T - T_i)$. Here, $\alpha_p = -1/\rho (\partial \rho/\partial T)_P$ is the thermal expansion coefficient $\alpha'_p$ normalised by $1/T'_c$. At the initial instant when $\vec{U} = 0$, the momentum equation Eq. (2) is reduced to the equation of hydrostatic equilibrium $\vec{\nabla} p_i = 1/Fr \, \rho_i \vec{g}$ which is subtracted from Eq. (2). The density difference $\rho - \rho_i$ is replaced with the term following from the simplified linear equation of state. The difference $p - p_i$ is denoted as $p_d$ where $p_d$ is the pressure component responsible only for motion. In the energy equation Eq. (7), the temperature $T$ substitutes by $T - T_i$ in the derivatives; note $T_i$ is a constant. The density is assumed to be constant everywhere except the buoyancy force. The model is applicable to low-speed flows ($M << 1$) where the contribution from $p_i$ and $p_d$ to the total pressure $P = \langle P \rangle + \gamma_0 M^2 (p_i + p_d)$ is very small therefore one can take $P \approx \langle P \rangle$. The volume-average pressure $\langle P \rangle$ is defined by the volume-average temperature $\langle T \rangle$. As a result, we have the Oberbeck-Boussinesq approximation

$$\nabla \cdot \vec{U} = 0 \qquad (17)$$

$$\frac{\rho_i}{\varphi} \frac{\partial \vec{U}}{\partial t} = -\vec{\nabla} p_d - \frac{1}{Fr} \alpha_p \rho_i (T - T_i) \vec{g} - \frac{1}{Da \operatorname{Re}} \frac{1}{K} \vec{U} \qquad (18)$$

$$\left((1-\varphi)\rho_s c_s + \varphi \rho_i c_p\right) \frac{\partial (T - T_i)}{\partial t} + \rho_i c_p (\vec{U} \cdot \nabla)(T - T_i) =$$
$$= -\varphi(\gamma_0 - 1) T \alpha_p \frac{d(\langle P \rangle - \langle P \rangle_i)}{dt} + \frac{\gamma_0}{\operatorname{RePr}_0} \lambda_m \Delta(T - T_i),$$

$$\lambda_m = 1 + \varphi \lambda_0 \Lambda (T-1)^{-\psi} \qquad (19)$$

$$\langle P \rangle - \langle P \rangle_i = \frac{\rho_i}{1 - \rho_i/3} \left(\langle T \rangle - \langle T \rangle_i\right), \quad \langle T \rangle = \int_\Omega T \, d\varpi \qquad (20)$$

In Eqs. (17)-(20), acoustic solutions are filtered as the equation of state Eq. (20) does not include the dynamic pressure $p_d$.

Because our study focuses on gravity-driven convection, it is worth to consider the Rayleigh-Darcy number $Ra_0$ defined as

$$Ra_0 = \frac{\Theta'_0 g' h' K'_0 \rho'^2_c (c'_{v0} + B')}{T'_c \lambda'_{m0} \eta'_0} \qquad (21)$$

The parameter $Ra_0$ is dependent on the other dimensionless parameters as

$$Ra_0 = \frac{\Theta \operatorname{Re}^2 Da \operatorname{Pr}_0}{Fr} \qquad (22)$$

and may be incorporated into Eq. (18) instead of

$Fr$.

Initially, the case of negligible stratification effects is considered. We find the conditions for convection of near-critical fluid has some similar features to convection of normal fluid. We compare steady-state regimes when a contribution from the piston effect in a near-critical fluid is reduced to zero. The momentum equation Eq. (18) including $Ra_0$ instead of $Fr$ and the energy equation Eq. (19) are simplified in a steady-state regime to the form

$$\vec{\nabla} p_d = -\frac{\alpha_p Ra_0}{\text{Re}^2 Da \text{Pr}_0} \rho_i \left(\frac{T-T_i}{\Theta}\right) \vec{g} - \frac{1}{Da\text{Re}} \frac{1}{K} \vec{U} \quad (23)$$

$$(\vec{U} \cdot \nabla)\left(\frac{T-T_i}{\Theta}\right) = \frac{\gamma_0 \lambda_m}{\text{Re} \text{Pr}_0 c_p} \frac{1}{\rho_i} \Delta\left(\frac{T-T_i}{\Theta}\right) \quad (24)$$

We write the variable $(T-T_i)/\Theta$ instead of $(T-T_i)$ in Eq. (23). Equation (24) is invariant of $(T-T_i)$ therefore one can write $(T-T_i)/\Theta$ instead of $(T-T_i)$ as well.

Approaching to the critical point, the Rayleigh-Darcy and Prandtl numbers have to grow but $Ra_0$ and $\text{Pr}_0$ stay constant. The numbers $Ra_0$ and $\text{Pr}_0$ are constructed of non-critical reference parameters whereas criticality is generated by the equation of state giving $\alpha_p$, $c_p$ those diverge at the critical point and by $\lambda_m$ diverging as well. Really, convection is characterized by some values $Ra$ and $\text{Pr}$ involving critical properties. The values $\text{Re}$, $Da$, $\gamma_0$ do not feel criticality and stay constant. The value $K$ is taken to be constant because we did not reveal any quantitative data on the near-critical behaviour of permeability. For comparison, we consider convection of normal fluid at other numbers $Ra$ and $\text{Pr}$. The fluid is assumed to be an ideal gas with the equation of state $P = \rho T$ and $\alpha_p = 1$, $c_p = \gamma_0$, $\lambda_m = 1$. Let $\Theta_{ig}$ be the temperature difference at boundaries in the ideal gas. Convection of ideal gas is described, instead of Eqs. (23), (24), by the equations:

$$\vec{\nabla} p_d = -\frac{Ra}{\text{Re}^2 Da\text{Pr}} \rho_i \left(\frac{T-T_i}{\Theta_{ig}}\right) \vec{g} - \frac{1}{Da\text{Re}} \frac{1}{K} \vec{U} \quad (25)$$

$$(\vec{U} \cdot \nabla)\left(\frac{T-T_i}{\Theta_{ig}}\right) = \frac{1}{\text{Re}\text{Pr}} \frac{1}{\rho_i} \Delta\left(\frac{T-T_i}{\Theta_{ig}}\right) \quad (26)$$

Convection of near-critical fluid and ideal gas will exhibit similar features, if the coefficients in analogous terms in Eqs. (23) and (25), (24) and (26) are equal to each other. We assume that terms $(T-T_i)/\Theta$ in Eqs. (23) and (24) are equal to $(T-T_i)/\Theta_{ig}$ in Eqs. (25) and (26). The condition of similarity leads to the relations $Ra_0 \alpha_p / \text{Pr}_0 = Ra/\text{Pr}$, $\gamma_0 \lambda_m /(\text{Pr}_0 c_p) = 1/\text{Pr}$ which give

$$Ra = Ra_0 \frac{\alpha_p c_p}{\gamma_0 \lambda_m}, \quad \text{Pr} = \text{Pr}_0 \frac{c_p}{\gamma_0 \lambda_m} \quad (27)$$

We want to satisfy Eqs. (27) so that the Froude number $Fr$ being the combinations of other parameters remains constant. Writing the expression of $Ra$ for the ideal gas by analogy with Eq. (22) as $Ra = \Theta_{ig} \text{Re}^2 Da\text{Pr}/Fr$ and combining the last expression with Eqs. (22), (27), one can obtain

$$\Theta_{ig} = \Theta \alpha_p \quad (28)$$

Following from the definition $\alpha_p = -1/\rho (\partial \rho/\partial T)_P$, thermodynamic identity for $c'_p$ [40] (normalized by $c'_{v0}$)

$$c_p = 1 - (\gamma_0 - 1)\frac{T}{\rho^2}\left(\frac{\partial P}{\partial T}\right)_\rho^2 \left(\frac{\partial \rho}{\partial T}\right)_P \quad (29)$$

and expression of $\lambda_m$ Eq. (19), the expressions of $Ra$ and $\text{Pr}$ Eqs. (27) give rise to the form

$$Ra = -Ra_0 \left(\frac{\partial \rho}{\partial T}\right)_P \left\{\frac{1}{\gamma_0} - \frac{(\gamma_0-1)}{\gamma_0} \frac{T}{\rho^2}\left(\frac{\partial P}{\partial T}\right)_\rho \left(\frac{\partial \rho}{\partial T}\right)_P\right\}$$
$$\times \left(1 + \varphi \lambda_0 \Lambda (T-1)^{-\psi}\right)^{-1} \quad (30)$$

$$\text{Pr} = \text{Pr}_0 \left\{ \frac{1}{\gamma_0} - \frac{(\gamma_0 - 1)}{\gamma_0} \frac{T}{\rho^2} \left( \frac{\partial P}{\partial T} \right)_\rho \left( \frac{\partial \rho}{\partial T} \right)_P \right\}$$

$$\times \left(1 + \varphi \lambda_0 \Lambda (T-1)^{-\psi}\right)^{-1} \quad (31)$$

called the correlation relations. The expression of $\Theta_{ig}$ Eq. (28) reduces to the equation

$$\Theta_{ig} = -\Theta \frac{1}{\rho} \left( \frac{\partial \rho}{\partial T} \right)_P \quad (32)$$

The relations Eqs. (30), (31) connect the model parameters $\text{Pr}_0$, $Ra_0$ (appearing in the governing equations) with the criteria of similarity $\text{Pr}$, $Ra$ (characterising convection actually). A dynamic behavior of near-critical fluid inside a porous medium at $\text{Pr}_0$, $Ra_0$ has similar features with that of ideal gas at $\text{Pr}$, $Ra$ in the governing equations. To achieve the required ratio between $Ra$ and $Ra_0$, the temperature difference at boundaries in a layer with the ideal gas $\Theta_{ig}$ should be taken instead of $\Theta$ according to Eq. (32). However the similarity is not full, as obtained numerically and will be discussed later. The same relations based on an intuitive conception were written in [30, 31]. The correlation relations for the Rayleigh and Prandtl numbers in a pure near-critical fluid were given in [21].

For the van der Waals gas, derivatives in Eqs. (30)-(32) can be easily found from the equation of state in the form of Eq. (11) as $(\partial \rho / \partial T)_P = (\partial m / \partial \varepsilon)_\pi$, $(\partial P / \partial T)_\rho = (\partial \pi / \partial \varepsilon)_m$. In this case, we get

$$Ra = \frac{2}{3} Ra_0 \varepsilon^{-1} \left( \frac{1}{\gamma_0} + \frac{\gamma_0 - 1}{\gamma_0} \frac{(1+\varepsilon)}{\varepsilon} \right) \left(1 + \varphi \lambda_0 \Lambda \varepsilon^{-\psi}\right)^{-1} \quad (33)$$

$$\text{Pr} = \text{Pr}_0 \left( \frac{1}{\gamma_0} + \frac{\gamma_0 - 1}{\gamma_0} \frac{(1+\varepsilon)}{\varepsilon} \right) \left(1 + \varphi \lambda_0 \Lambda \varepsilon^{-\psi}\right)^{-1} \quad (34)$$

$$\Theta_{ig} = \frac{2}{3\varepsilon} \Theta \quad (35)$$

The relations Eqs. (33), (34) show how the real criteria of similarity depend on the temperature distance to the critical point defined by $\varepsilon$. Approaching to the critical point ($\varepsilon \to 0$), $Ra$ and $\text{Pr}$ grow unboundedly: $Ra \sim \varepsilon^{\psi - 2} \to \infty$, $\text{Pr} \sim \varepsilon^{\psi - 1} \to \infty$ ($\psi < 1$ in the most real cases) according to expectations. However, the indexes $\psi - 2$ and $\psi - 1$ in the asymptotic dependences are not accurate since they follow from the van der Waals equation of state. The temperature difference $\Theta_{ig} \sim \varepsilon^{-1}$ has to grow unboundedly as well.

Now a stratified fluid phase is considered. Taking into account the Schwarzschild criterion, the Rayleigh-Darcy number should be corrected for the contribution from the adiabatic temperature gradient $\Gamma'_a$. The linear analysis [41] shows that the applied temperature gradient $\Gamma'$ should substitute by the difference $\Gamma' - \Gamma'_a$; this correction is similar to that in pure compressible fluids [6]. As a result, the Rayleigh-Darcy number is scaled by the stratification coefficient $k = 1 - \Gamma'_a / \Gamma'$. If $k \le 0$, the Rayleigh-Darcy number is taken to be zero. In the general case, $\Gamma'_a = g' \alpha'_p T' / c'_p$ and, in the considered problem, $\Gamma' = \Theta' / h'$ leading to the relation at $k > 0$

$$Ra^s = Ra\, k, \quad k = 1 - \frac{g' \alpha'_p T' h'}{c'_p \Theta'} \quad (36)$$

Here, $Ra^s$ is the corrected real Rayleigh-Darcy number taking into account the effect of density stratification. We can substitute $\alpha'_p$ and $c'_p$ in $k$ by their thermodynamic identities and refer to the equation of state. In the case of the van der Waals gas, Eqs. (36) yield the form

$$Ra^s = Ra\, k, \quad k = 1 - \frac{\Theta_{ac}(\gamma_0 - 1)(1+\varepsilon)}{\Theta(\varepsilon + (\gamma_0 - 1)(1+\varepsilon))},$$

$$\Theta_{ac} = \frac{2M^2 \gamma_0}{3 Fr} \quad (37)$$

The coefficient $k$ changes from $k = 1 - \Theta_{ac}(\gamma_0 - 1)/(\Theta \gamma_0)$ at $\varepsilon \gg 1$ to $k = 1 - \Theta_{ac}/\Theta$ at $\varepsilon \to 0$. Here, $\Theta_{ac}$ is the temperature difference at boundaries defined as the product of the adiabatic temperature gradient at the critical point and the width of layer. If $\Theta_{ac}/\Theta \ll 1$, one obtains $k \approx 1$ independently of approach to the critical point that is the stratification is of little importance. If $\Theta_{ac}/\Theta \geq \gamma_0/(\gamma_0-1)$, we have $k \leq 0$ and $Ra^s = 0$ that means the stabilizing effect due to stratification. It is even possible, if $1 < \Theta_{ac}/\Theta < \gamma_0/(\gamma_0-1)$, that going to the critical point ($\varepsilon \to 0$), the coefficient $k$ being initially positive becomes smaller, then negative that corresponds to convection to be totally suppressed due to stratification. Suppressing of convection in a pure near-critical fluid in a close critical vicinity was obtained by numerical simulations [19].

Under the influence of stratification, the similarity of convection of near-critical and normal fluid phases is expected to be achieved with the equality of the corrected real Rayleigh-Darcy numbers $Ra^s$ including the stratification effect. It gives the condition $Ra_{ig}k_{ig} = Ra\,k$, where $Ra_{ig}$ and $k_{ig}$ are the Rayleigh-Darcy number disregarding stratification and the stratification coefficient in the ideal gas. Following from the general expression, one can obtain $k_{ig} = 1 - g'h'/((c'_{v0}+B')\Theta'_{ig})$ that leads to the form

$$Ra_{ig} = Ra\frac{k}{k_{ig}},\ k_{ig} = 1 - \frac{\Theta_{iga}}{\Theta_{ig}},$$

$$\Theta_{iga} = \frac{M^2(\gamma_0-1)}{Fr} \quad (38)$$

Here, $\Theta_{iga}$ is the temperature difference at boundaries defined as the adiabatic temperature gradient in the ideal gas multiplied by the width of layer. One can find $\Theta_{ig}$ following from Eqs. (38), the correlation relations and some obvious dependencies. We find

$$\Theta_{ig} = \Theta_{iga} + \frac{2}{3}\varepsilon^{-1}\left(\Theta - \frac{2\Theta_{iga}\gamma_0(1+\varepsilon)}{3(\varepsilon+(\gamma_0-1)(1+\varepsilon))}\right) \quad (39)$$

The expression of $\Theta_{ig}$ in the stratified ideal gas Eq. (39) includes additional terms as compared with Eq. (35) defined under the negligible effect of stratification.

Remember, some similarity of convection of near-critical and normal fluid phases can be achieved only in steady-state regimes when, in the first case, the contribution from the piston effect is reduced to zero. In unsteady regimes, comparing such two media, one can realize changes in convective structure and heat transfer due to the piston effect.

### III. RESULTS OF NUMERICAL SIMULATIONS

An infinite horizontal porous layer at the width $h' = 0.1$ m formed by a sand-like matter at $c'_s = 960$ J/(kg K), $\rho'_s = 2400$ Kg/m$^3$, $\lambda'_s = 3.35$ W/(m K), and $K'_0 = 4 \times 10^{-11}$ m$^2$ under the Earth's gravity is considered. The layer is saturated with carbon dioxide having the critical parameters: $T'_c = 304.15$ K, $\rho'_c = 4.68 \times 10^2$ kg/m$^3$, $P'_c = 7.387$ MPa [5], and the other parameters: $\eta'_0 = 32.57 \times 10^{-6}$ Pa s [42], $c'_{v0} = 567$ J/(kg K), $\lambda'_0 = 0.05$ W/(m K), $\Lambda = 2.8 \cdot 10^{-2}$, $\psi = 0.74$. The value $c'_{v0}$ is the classical heat capacity of multi-atomic gas normalized by the molecular weight, $c'_{v0} = 3R'/\mu'_g$; here $\mu'_g = 0.044$ kg/mol. The values $\lambda'_0$, $\Lambda$, $\psi$ are defined by the experimental data on the thermal conductivity of $CO_2$ near the critical point [43]. The reference velocity is $U' = 0.276$ m/s; $U'$ is not the characteristic velocity of considered dynamic phenomena and is used only as a scale parameter. The dimensionless parameters are: $Fr = 7.80 \times 10^{-2}$, $Pr_0 = 1.21 \times 10^{-2}$, $Re = 3.97 \times 10^5$, $M = 1 \times 10^{-3}$, $Da = 4 \times 10^{-9}$, $\gamma_0 = 1.33$, $\rho_s c_s = 8.68$, $\lambda_0 = 2.46 \times 10^{-2}$.

The fluid is stratified but the average density is critical. The temperature is slightly above the critical one defined by $\varepsilon_i = (T'_i - T'_c)/T'_c$. In the initial instant, the temperature at the bottom boundary grows in $\Theta$. The top temperature is held constant. We study steady-state convection and perform integration at the variable time step until a steady-state regime is reached. The idealized situation when the permeability (and, consequently, $Da$) does not depend on $\varphi$ is considered. The values $\varepsilon_i$ and $\Theta$ are chosen so that the Oberbeck-Boussinesq approximation is applicable although the solver is based on the full model Eqs. (1)-(5) and allows one to solve non-Boussinesq problems as well. We are interested in solutions close to the convection threshold where the two-dimensional roll structure occurs [44]. A square region at the aspect ratio $l'/h'=1$ containing a single convective roll is considered. Periodic conditions at the vertical boundaries are applied. Sketch of the problem is shown in Fig. 1.

To characterize the convective heat transfer, the Nusselt number $Nu$ at the bottom boundary is defined by the expression

$$Nu = \frac{1}{j'}\int_0^{l'}\lambda'_m \frac{\partial T'}{\partial y'}dx' \qquad (40)$$

where $j'$ is the heat flux due to the thermal conductivity, $j' = \lambda'_m \Theta'/h'$. In simulations, the magnitudes of $Nu$ at the top and bottom were equal to each other.

Before simulations, we examined grid effects. The problem at $\varphi = 0.7$, $\varepsilon_i = 2.72\times10^{-4}$, and $\Theta = 1.1607\times10^{-5}$ was solved using 41x41, 61x61, 81x81, and 101x101 uniform grids. For these cases, computations give $Nu - 1 = 3.92\times10^{-3}$, $3.95\times10^{-3}$, $3.96\times10^{-3}$, and $3.96\times10^{-3}$, respectively. All demonstrated results are obtained with the use of 81x81 grid as a more fine spacing does not effect on the numerical solution.

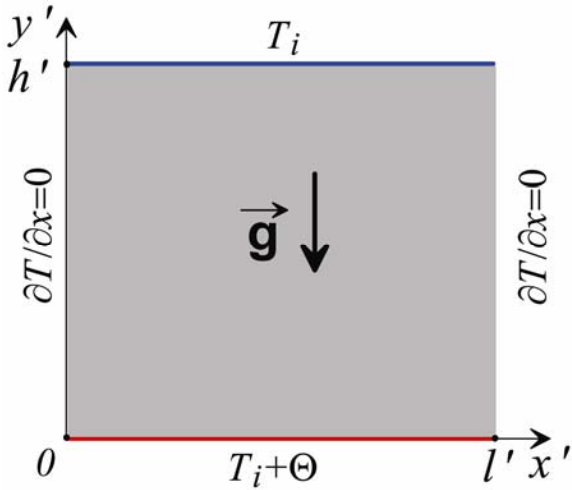

**FIG. 1**. Sketch of the Rayleigh-Darcy problem.

Six series of solutions at different $\varphi$ and $\varepsilon_i$ were obtained; $\Theta$ varies in all series to change the distance to the stability threshold. The values of $\varepsilon_i$ are chosen so that to get (i) a not severe ($\varepsilon_i = 5.43\times10^{-3}$, $k = 0.870 \div 0.876$), (ii) intermediate ($\varepsilon_i = 1.81\times10^{-3}$, $k = 0.432\div0.454$), and (iii) dominant ($\varepsilon_i = 2.72\times10^{-4}$, $k = 0.0185\div0.0215$) role of Schwarzschild criterion. The variations in $\Theta$ are: (i) $8.6\times10^{-5}\div9.0\times10^{-5}$, (ii) $1.19\times10^{-5}\div2.07\times10^{-5}$, and (iii) $1.1572\times10^{-5}\div1.1607\times10^{-5}$. The variations in $Ra_0$ calculated by Eq. (22) are: (i) $8.41\times10^{-3}\div8.802\times10^{-3}$, (ii) $1.946\times10^{-3}\div2.024\times10^{-3}$, and (iii) $1.1317\times10^{-3}\div1.1352\times10^{-3}$. Convection of near-critical fluid is characterized by the real Rayleigh-Darcy number $Ra^s$ which is calculated by Eqs. (33) and (37); we treat $\varepsilon$ as $\varepsilon_i + 0.5\Theta$ being the average of the temperature at the top and bottom. The magnitudes of $Nu$ are obtained numerically. The points $Nu(Ra^s)$ are plotted in Fig. 2(a) and compared with the theoretical solution for a porous layer filled with a low compressible fluid phase given by the stability analysis in the form [44]

$$Nu = 1 + 2(1 - Ra^*/Ra^s) \qquad (41)$$

Equation (41) is valid at $4\pi^2 \leq Ra^s \leq 16\pi^2$. Here, $Ra^* = 4\pi^2 = 39.48$ is the threshold Rayleigh-Darcy number corresponding to the convection onset. As clear, the points $Nu(Ra^s)$ in simulations situate below the theoretical curve; the deflection from the curve becomes larger with decreasing in $\varepsilon_i$ (being independent of $\varphi$) because the stratification effect enlarges.

As discussed in [13, 22], considering the convective heat transfer in stratified fluids, the Nusselt number has to be corrected for the contribution from the adiabatic temperature gradient. In our notations, the corrected Nusselt number is

$$Nu^s = \frac{1}{j'_s} \int_0^{l'} \lambda'_m \left( \frac{\partial T'}{\partial y'} - \Gamma'_a \right) dx' \qquad (42)$$

where $j'_s = \lambda'_m (\Theta'/h' - \Gamma'_a)$ is the corrected heat flux due to the thermal conductivity. One can obtain the following relation between $Nu^s$ and $Nu$

$$Nu^s = 1 + (Nu - 1)/k \qquad (43)$$

where $k$ is defined by Eq. (37). We calculated $Nu^s$

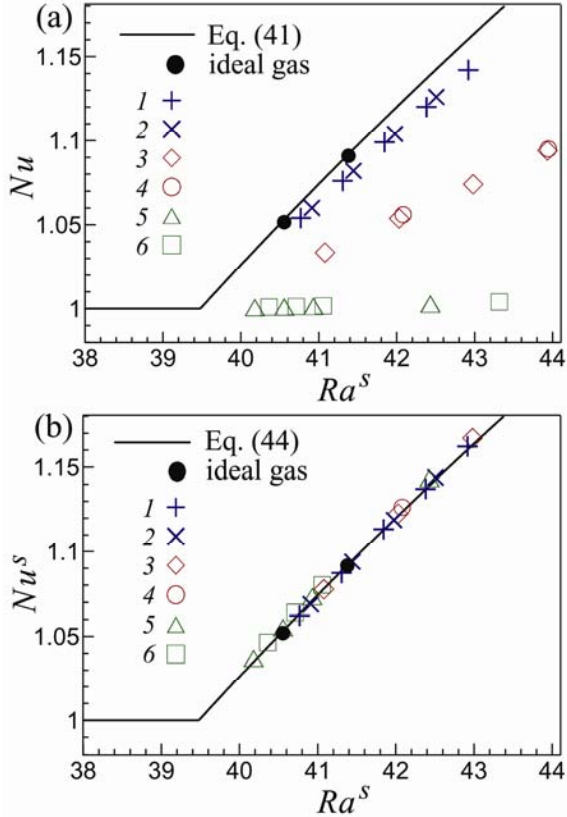

**FIG. 2.** The Nusselt numbers $Nu$ (a) and $Nu^s$ (b) versus the Rayleigh-Darcy number $Ra^s$ in a layer with a near-critical fluid at $\varepsilon_i$=5.43x10$^{-3}$ and $\varphi$=0.4 (1), 0.7 (2); $\varepsilon_i$=1.81x10$^{-3}$ and $\varphi$=0.4 (3), 0.7 (4); $\varepsilon_i$=2.72x10$^{-4}$ and $\varphi$=0.4 (5), 0.7 (6). Solid lines: analytical dependencies Eqs. (41) (a) and (44) (b). Solid circles: a layer with an ideal gas.

by Eq. (43) and plotted the points $Nu^s(Ra^s)$ in Fig. 2(b). As clear, these points hit on the curve precisely demonstrating that the relation

$$Nu^s = 1 + 2(1 - Ra^*/Ra^s) \qquad (44)$$

is valid under considered conditions. Data in Fig. 2(b) indicate an analogy of heat transfer of high compressible near-critical and low compressible normal fluid phases. To obtain the results in Fig. 2, numerical simulations have to be performed at a very high accuracy because, with approach to the critical point, the fluid phase becomes very sensitive to the temperature disturbances. For example, at $\varepsilon_i = 2.72 \times 10^{-4}$ the change in $\Theta$ in 0.02 % induces the change in $Nu^s$ in 0.9 %.

Note, second condition Eqs. (16) for the Oberbeck-Boussinesq approximation to be applicable may not be satisfied very close to the critical point. This condition reduced at $\varepsilon \ll 1$ to the form $\varepsilon \gg \left(2\gamma_0 M^2/(3Fr)\right)^{1/(\beta+\gamma)}$ gives at the taken parameters $\varepsilon \gg 1.96 \times 10^{-4}$ which is not valid, for example, at $\varepsilon = 2.72 \times 10^{-4}$. However, as shown in Fig. 2(b), the results of simulation at $\varepsilon_i = 2.72 \times 10^{-4}$ are in a good agreement with the theoretical prediction based on the Oberbeck-Boussinesq approximation.

Convection in a porous layer filled with an ideal gas to obey the equation of state Eq. (10) at $a = b = 0$ was also simulated numerically. The parameters are: $\Pr_0 = 9.678$ (corresponding to $\Pr$ in a layer with a near-critical fluid at $\varepsilon_i = 2.72 \times 10^{-4}$ and $\varphi = 0.4$), and $\Theta_{ig} = 5.227 \times 10^{-4}$, $5.332 \times 10^{-4}$. The other parameters are as in the near-critical layer above. The stratification coefficient calculated by Eq. (38) is $k_{ig} = 0.9919$, 0.9921 for two taken magnitudes of $\Theta_{ig}$, respectively. The magnitudes of $Nu^s$ are calculated by Eq. (43) which is used at $k = k_{ig}$. In Fig. 2(b), the points $Nu^s(Ra^s)$ for the ideal gas lie on the theoretical curve according to expectations.

To compare convection of near-critical and normal fluids in detail, Fig. 3 exhibits the distribution of different variables at $Pr = 9.678$, $Ra^s = 40.56$ and $\varphi = 0.4$. In the layer with the near-critical fluid: $\varepsilon_i = 2.72 \times 10^{-4}$, $\Theta = 1.1574 \times 10^{-5}$ and ideal gas: $\Theta_{ig} = 5.227 \times 10^{-4}$; other parameters are as above. As clear, the temperature, density and velocity variables in Figs. 3(a, b) behave similarly in two cases. However $(T - T_i)/\Theta$ in the ideal gas varies along the horizontal in a much wider range than in the near-critical fluid (Fig. 3(c)). Figure 4 confirms that the density patterns are identical whereas the isotherms in the near-critical fluid phase are distorted less. This tendency is explained by very different thermal expansion coefficients $\alpha_p$ in two fluids. In a near-critical fluid, $\alpha_p$ is very large leading to the significant density variations induced by small temperature disturbances.

The simulated points $Nu^s(Ra^s)$ hit on the theoretical curve only near the convection onset at $(Ra^s - Ra^*)/Ra^* \leq 0.1$. At larger $Ra^s$, the simulated values of $Nu^s$ are in systematic excess of the theoretical prediction resulted from a restriction on the horizontal size of layer. In theory, the layer is unlimited leading to that the wavenumber of perturbations in the stability analysis depends

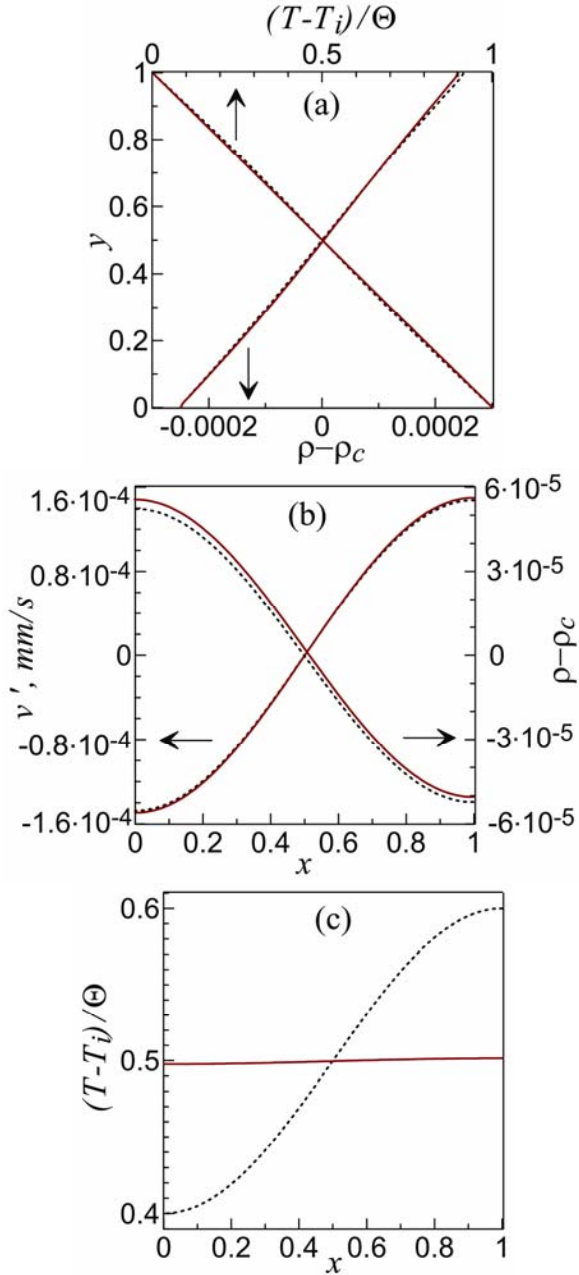

**FIG. 3.** The density $\rho - \rho_c$ and the temperature $(T-T_i)/\Theta$ along the middle vertical line (a), the vertical component of the velocity $v'$, the density $\rho - \rho_c$ (b), and the temperature $(T-T_i)/\Theta$ (c) along the middle horizontal line. Solid lines: a layer with a near-critical fluid. Broken lines: a layer with an ideal gas; the subscript "ig" at $\Theta$ in an ideal gas is avoided.

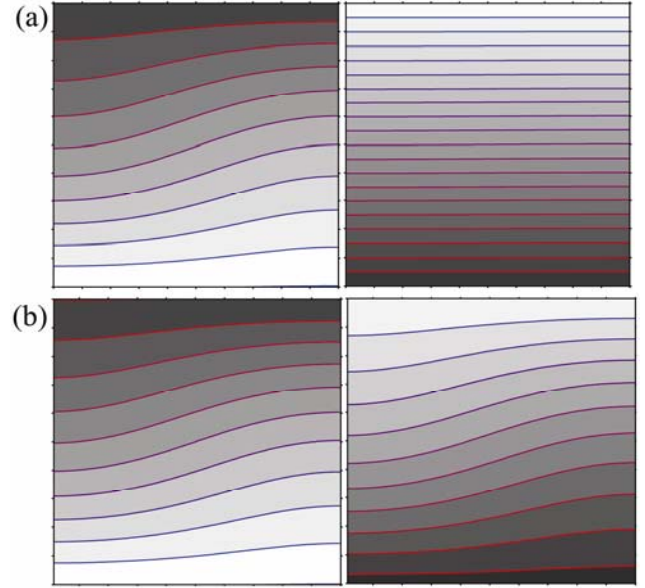

**FIG. 4.** The density (left) and the temperature (right) patterns in a layer with a near-critical fluid (a) and an ideal gas (b). The range of density $\rho - \rho_c$: -2.5x10$^{-4}$ ÷ 2.5x10$^{-4}$. The range of temperature $(T-T_i)/\Theta$: 0 ÷ 1.

continuously on $Ra^S$ whereas, in our simulations, the wavenumber stays constant in a square cell.

The analysis of steady convective heat transfer in a pure stratified near-critical $^3$He heated from below was fulfilled in [13] to compare experimental and numerical data in a wide range of the Rayleigh number. The convection heat current versus the corrected Rayleigh number collapses in general onto a single curve as our data in Fig. 2(b).

## IV. ANALYSIS OF THE CONVECTION ONSET

As shown in Fig. 2(b), the convection onset is defined by $Ra^S$ which have to be equal to $Ra^*$ independently of $\varepsilon_i$ and $\varphi$. This result allows one to find the temperature difference at boundaries $\Theta^*$ corresponding to the convection onset; superscript "*" denotes the threshold values. Combining Eqs. (22), (33), (37), we get the relation

$$Ra^S = \frac{2}{3}\frac{\Theta \mathrm{Re}^2 Da \Pr_0}{Fr}\varepsilon^{-1}\left(\frac{1}{\gamma_0}+\frac{\gamma_0-1}{\gamma_0}\frac{(1+\varepsilon)}{\varepsilon}\right)$$
$$\times\left(1+\varphi\lambda_0\Lambda\varepsilon^{-\psi}\right)^{-1}\left(1-\frac{\Theta_{ac}(\gamma_0-1)(1+\varepsilon)}{\Theta(\varepsilon+(\gamma_0-1)(1+\varepsilon))}\right) \quad (45)$$

which can be used in the threshold conditions ($Ra^S$ and $\Theta$ are replaced with $Ra^*$ and $\Theta^*$, respectively)

$$Ra^* = \frac{2}{3}\frac{\Theta^* \mathrm{Re}^2 Da \Pr_0}{Fr}\varepsilon^{-1}\left(\frac{1}{\gamma_0}+\frac{\gamma_0-1}{\gamma_0}\frac{(1+\varepsilon)}{\varepsilon}\right)$$
$$\times\left(1+\varphi\lambda_0\Lambda\varepsilon^{-\psi}\right)^{-1}\left(1-\frac{\Theta_{ac}(\gamma_0-1)(1+\varepsilon)}{\Theta^*(\varepsilon+(\gamma_0-1)(1+\varepsilon))}\right) \quad (46)$$

The threshold stratification coefficient $k^*$, following from Eq. (37), is

$$k^*=1-\frac{\Theta_{ac}(\gamma_0-1)(1+\varepsilon)}{\Theta^*(\varepsilon+(\gamma_0-1)(1+\varepsilon))} \quad (47)$$

The equation Eq. (46) gives rise to the explicit expression of $\Theta^*$:

$$\Theta^* = \Theta_{Ra}^* + \Theta_{Sc}^* \quad (48)$$

$$\Theta_{Ra}^* = \frac{3Ra^* Fr}{2\mathrm{Re}^2 Da \Pr_0}\varepsilon\left(1+\varphi\lambda_0\Lambda\varepsilon^{-\psi}\right)$$
$$\times\left(\frac{1}{\gamma_0}+\frac{\gamma_0-1}{\gamma_0}\frac{(1+\varepsilon)}{\varepsilon}\right)^{-1} \quad (49)$$

$$\Theta_{Sc}^* = \frac{\Theta_{ac}(\gamma_0-1)(1+\varepsilon)}{(\varepsilon+(\gamma_0-1)(1+\varepsilon))} \quad (50)$$

We separate two terms $\Theta_{Ra}^*$ and $\Theta_{Sc}^*$ in $\Theta^*$. The term $\Theta_{Ra}^*$ is obtained from Eq. (46) at $k^*=1$ (the stratification is negligible) and associated with the Rayleigh criterion. The term $\Theta_{Sc}^*$ is obtained from Eq. (47) at $k^*=0$ (the applied and adiabatic temperature gradients are equal to each other) and attributed to the Schwarzschild criterion.

One can find that, far away from the critical point ($\varepsilon \gg 1$), $\Theta_{Ra}^*$ grows with an increase in $\varepsilon$ as $\Theta_{Ra}^* \approx \varepsilon$ but $\Theta_{Sc}^*$ goes to the limit: $\Theta_{Sc}^* \to \Theta_{ac}(\gamma_0-1)/\gamma_0$. These tendencies lead to $\Theta_{Ra}^* \gg \Theta_{Sc}^*$ that is the Rayleigh criterion dominates in determining the threshold temperature difference $\Theta^*$. As the distance to the critical point decreases ($\varepsilon \to 0$), $\Theta_{Ra}^*$ approaches to zero as $\Theta_{Ra}^* \sim \varepsilon^{2-\psi}$ and $\Theta_{Sc}^*$ goes to the other finite limit: $\Theta_{Sc}^* \to \Theta_{ac}$ resulting in $\Theta_{Sc}^* \gg \Theta_{Ra}^*$ and the dominant role of the Schwarzschild criterion.

The dependencies of $\Theta_{Ra}^*$, $\Theta_{Sc}^*$, and $\Theta^*$ on $\varepsilon$ calculated by Eqs. (48)-(50) are exhibited in Fig. 5(a). We consider the physically realistic case when a porous medium is formed by spheres of equal diameters of identical matter. In this case, the permeability $Da$ depends on the porosity $\varphi$ according to the analytical relation $Da(1-\varphi)^2/\varphi^3 = const$ [44]. We take $\varphi = 0.4, 0.7$ and $Da = 4.00 \times 10^{-9}$, $6.48 \times 10^{-6}$, respectively, to satisfy this relation; other parameters are as above. The curves 1 and 4 in Fig. 5(a) coincide with each other. It is shown, that at some distance from the critical point where the Rayleigh criterion is major ($\Theta^* \to \Theta_{Ra}^*$), the value $\Theta^*$ increases if $\varphi$ decreases (the fraction of solid phase grows). It means that the solid phase stabilizes the state of fluid due to a reduction in the permeability. In the close vicinity of the critical point where the Schwarzschild criterion is dominant ($\Theta^* \to \Theta_{Sc}^*$), one can obtain by Eq. (50) that $\Theta^* \to \Theta_{ac}$. It means that $\Theta^*$ becomes independent of solid phase and is determined solely by the adiabatic temperature gradient of fluid phase at the critical point.

Using Eqs. (49), (50), one can simplify the expressions of $k$ Eq. (37) and $Ra^s$ Eq. (45) to the form

$$k = 1 - \frac{\Theta_{Sc}^*}{\Theta}, \quad Ra^s = Ra^* \frac{\Theta - \Theta_{Sc}^*}{\Theta_{Ra}^*} \quad (51)$$

The threshold coefficient $k^*$ may be written instead of Eq. (47) as

$$k^* = 1 - \frac{\Theta_{Sc}^*}{\Theta_{Ra}^* + \Theta_{Sc}^*} \quad (52)$$

In Fig. 5(b), $k^*$ versus $\varepsilon$ calculated by Eq. (52) at $\Theta_{Ra}^*$ and $\Theta_{Sc}^*$ defined by Eqs.(49), (50) is drawn. As clear, at $\varepsilon \approx 1$ where $\Theta_{Ra}^* \gg \Theta_{Sc}^*$, $k^* \to 1$. At $\varepsilon \leq 1 \times 10^{-4}$ where $\Theta_{Sc}^* \gg \Theta_{Ra}^*$, $k^* \to 0$. In the intermediate range of $\varepsilon$, an increase in the fraction of solid phase (corresponding to a decrease in $\varphi$) gives rise to the increase in $k^*$. Varying the porosity $\varphi$, one can control whether the Rayleigh or Schwarzschild criterion is major.

One can define changes in $Ra^s$ and $Nu^s$ depending on $\Theta$ at the stability threshold. The relation Eq. (44) is valid close to the convection onset and is used here. One obtains $\partial Ra^s / \partial \Theta \big|_{\Theta = \Theta^*}$ from Eq. (45) and $\partial Nu^s / \partial \Theta \big|_{\Theta = \Theta^*} = 2/Ra^* \, \partial Ra^s / \partial \Theta \big|_{\Theta = \Theta^*}$ from Eq. (44). The resultant expressions are

$$\left.\frac{\partial Ra^s}{\partial \Theta}\right|_{\Theta = \Theta^*} = \frac{Ra^*}{\Theta_{Ra}^*}, \quad \left.\frac{\partial Nu^s}{\partial \Theta}\right|_{\Theta = \Theta^*} = \frac{2}{\Theta_{Ra}^*} \quad (53)$$

Note, the derivatives $\partial Ra^s / \partial \Theta \big|_{\Theta = \Theta^*}$ and $\partial Nu^s / \partial \Theta \big|_{\Theta = \Theta^*}$ are determined only by the Rayleigh temperature difference $\Theta_{Ra}^*$. These dependencies look puzzling in the close vicinity of the critical point where the role of the Rayleigh criterion is negligible. Going to the critical point ($\varepsilon \to 0$), $\Theta_{Ra}^*$ approaches to zero (see Fig. 5(a)) and, consequently, $\partial Ra^s / \partial \Theta \big|_{\Theta = \Theta^*} \to \infty$ and $\partial Nu^s / \partial \Theta \big|_{\Theta = \Theta^*} \to \infty$ demonstrating that the fluid phase becomes very sensitive to the temperature disturbances.

## V. RESULTS AND CONCLUSIONS

We considered Rayleigh-Darcy convection in a porous layer filled with a near-critical fluid. With approach to the critical point, some thermodynamic coefficients of fluid phase grow unboundedly but the reference values in the Rayleigh-Darcy $Ra_0$ and Prandtl $Pr_0$ numbers in the nondimensional governing equations stay constant. Therefore, the numbers $Ra_0$ and $Pr_0$ called the model ones do not feel criticality and do not characterize convection actually. In the conditions of the Oberbeck-Boussinesq approximation to be valid, the correlation relations connecting the model $Ra_0$ (and $Pr_0$) and real $Ra^s$ (and $Pr$) Rayleigh-Darcy (and Prandtl) numbers in the case of a fluid phase subjected to the density stratification were deduced. The numbers $Ra^s$ and $Pr$ are the criteria of similarity. They take into account the real near-critical properties and characterize convection actually. The van der Waals gas was under study, however we gave a guideline for obtaining the correlation relations for different two-parameter equations of state. To characterize the convective heat transfer in a stratified fluid phase, the Nusselt number $Nu^s$ taking into account the stratification effect and including the difference between the applied and adiabatic temperature gradients was analyzed. Numerical simulations of steady-state convection slightly above the stability threshold showed that the dependence of $Nu^s$ on $Ra^s$ is universal and agrees with the theoretical predictions for a porous layer filled with a normal low compressible fluid. This result, on one hand, validates our correlation relations, on the other hand, gives the analytical relation $Nu^s(Ra^s)$ for the layer with a near-critical fluid phase. A comparison of numerical results on convection of near-critical and normal fluids at equal $Ra^s$ and $Pr$ exhibited that the similarity in convective heat transfer in two cases comes from very different temperature fields but like velocity and density patterns.

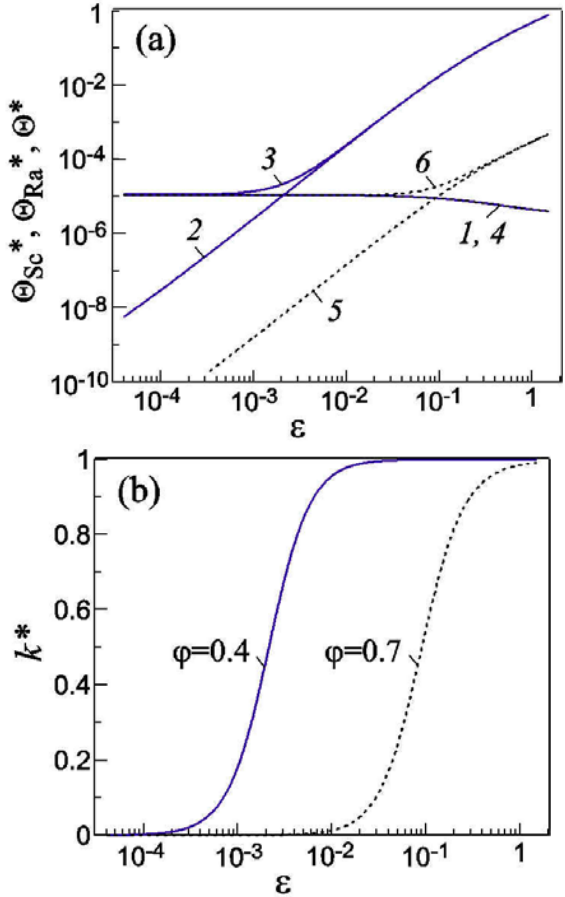

**FIG. 5.** The temperature differences at boundaries $\Theta_{Sc}^*$ (1, 4), $\Theta_{Ra}^*$ (2, 5) and $\Theta^*$ (3, 6) (a), and the stratification coefficient $k^*$ (b) at the porosity $\varphi=0.4$ (solid lines), 0.7 (broken lines).

We analyzed an influence of the Rayleigh and Schwarzschild criteria on the convection onset in terms of the threshold temperature difference at boundaries controlled by the first and second criteria, $\Theta_{Ra}*$ and $\Theta_{Sc}*$, respectively. Analytical expressions of $\Theta_{Ra}*$ and $\Theta_{Sc}*$ were obtained. The applied temperature difference $\Theta*$ changes from $\Theta_{Ra}*$ far away from the critical point to $\Theta_{Sc}*$ in a close critical vicinity. Near the critical point, $\Theta*$ becomes independent of solid matrix and is determined solely by the critical adiabatic temperature gradient of fluid phase. We found the stratification coefficient $k*$ as a function of $\Theta_{Ra}*$ and $\Theta_{Sc}*$ and the derivatives $\partial Ra^s / \partial \Theta \big|_{\Theta = \Theta^*}$ and $\partial Nu^s / \partial \Theta \big|_{\Theta = \Theta^*}$ those depend only on $\Theta_{Ra}*$. The last derivatives diverge at the critical point demonstrating a very high sensitivity of near-critical fluid phase to the temperature disturbances.

We did not give here any unsteady non-Boussinesq solutions however the solver employed allows us to do that. Non-Boussinesq effects become actual very close to the critical point. Really, if $\varepsilon \to 0$, the threshold temperature difference at boundaries $\Theta*$ tends to the finite limit (see Fig. 5(a)) that is the condition $\Theta* \ll \varepsilon$ for the Oberbeck-Boussinesq approximation to be valid is not satisfied in a close critical vicinity. We shall continue our study and give examples of non-Boussinesq behavior of near-critical fluid phase inside a porous matrix elsewhere.

## ACKNOWLEDGMENTS

The author is indebted to Professor V.I. Polezhaev for helpful discussions. This work received financial support from the Russian Foundation for Basic Research (Grant No. 09-01-00117).